\documentclass[fleqn,usenatbib]{mnras}

\usepackage{newtxtext,newtxmath}

\usepackage[T1]{fontenc}
\usepackage{subcaption}
\usepackage{times}             
\usepackage{natbib}
\DeclareRobustCommand{\VAN}[3]{#2}
\let\VANthebibliography\thebibliography
\def\thebibliography{\DeclareRobustCommand{\VAN}[3]{##3}\VANthebibliography}

\usepackage{graphicx}	
\usepackage{tablefootnote}
\usepackage{multirow}
\usepackage{multicol}
\usepackage{amsmath}	
\usepackage{float}

\newcommand{\gt}{>}

\title[Dynamically derived morphology]{Dynamically derived morphology from the recurrence patterns of close binary stars using Kepler data}
\author[A. R. V. Kashyap et al.]{
Anisha R. V. Kashyap,$^{1}$
D. Pawar,$^{2}$
R. Misra,$^{1}$
G. Ambika,$^{3}$
and Sandip V George$^{4}$\thanks{E-mail: sandip.george@abdn.ac.uk}
\\
$^{1}$Inter University Centre for Astronomy and Astrophysics, Pune - 411007, India\\
$^{2}$ Ramniranjan Jhunjhunwala College, Ghatkopar West, Mumbai-400086, India\\
$^{3}$Indian Institute of Science Education and Research Thiruvananthapuram, Thiruvananthapuram-695551, India\\
$^{4}$ University of Aberdeen, King's College, Aberdeen, AB24 3FX, United Kingdom
}

\date{Accepted XXX. Received YYY; in original form ZZZ}

\pubyear{2025}

\begin{document}
\label{firstpage}
\pagerange{\pageref{firstpage}--\pageref{lastpage}}
\maketitle

\begin{abstract}
We propose a novel method for classifying close binary stars based on the dynamical structure inherent in their light curves.  We apply this method to the light curves of binaries from the revised  Kepler Eclipsing  binary catalog, selecting close binaries which have the standard morphology parameter, $c$, $\gt 0.5$ corresponding to semi-detached, over-contact and ellipsoidal binary stars. Using the method of time delay embedding, we recreate the nonlinear dynamics underlying the data and quantify the patterns of recurrences in them. Using two of the recurrence measures, Determinism  and Entropy, we define a new Dynamically Derived Morphology (DDM) parameter and compute its values for close binary stars. While, as expected, this metric is somewhat inversely correlated with the existing morphology parameter (Spearman $\rho= -0.21$), the method offers an alternate classification scheme for the close binary stars that captures their nonlinear dynamics, an aspect often overlooked in conventional methods. Hence, the DDM parameter is expected to distinguish between stars with similar folded light curves, but are dynamically dissimilar due to nonlinear effects. Moreover, since the method can be easily automated and is computationally efficient, it can be effectively used for sensitive large data sets in future.
\end{abstract}




\section{Introduction}

\label{sect:intro}
Binary stellar systems consist of two stars orbiting a common center of mass bound by their mutual gravitational attraction~\citep{Shu1982book}.
When the component stars eclipse each other from our line of sight, resulting in observable dips in brightness, they are called eclipsing binaries (EBs). Morphologically, these stars are classified into detached (D), semi-detached (SD) or overcontact (OC) stars depending on the amount of mass and energy transfer between the component stars, which is determined by the extent to which their Roche lobes are filled~\citep{SHORE200377}. 
The latter two, together with ellipsoidal binaries (ELV), where tidal deformations also contribute to variability, are termed as close binary stars. 
The strong interactions between the components in close binaries result in deviations from periodicity causing unequal maxima and varying eclipse times, which are indicative of rich nonlinear dynamics in such stars~\citep{milone1968peculiar,knote2022characteristics,tran2013anticorrelated, George2020a}. 

Determining the morphology of eclipsing binary stars from their light curves is a major challenge in time domain astronomy. This is particularly true in the context of enhanced sensitivity of modern space observatories such as Gaia, Kepler and TESS, resulting in very large datasets of light curves. While this has opened up a new arena for the detection and study of stellar variability, the volume of data recorded is too large for traditional methods of classification to be efficient. Automated techniques based on neural networks \citep{Sarro2006a, Paegert2014a}, functional principal component analysis \citep{Modak2022a} and locally linear embedding\citep{Matijevi2012} have been proposed to overcome this issue. \textcolor{red}{For example, \citet{vcokina2021automatic} demonstrated that deep-learning models trained on synthetic light curves generated with ELISa\citep{vcokina2021elisa} can accurately distinguish detached and overcontact binaries, and \citet{Prsa2011a} used neural networks trained on PHOEBE-generated models to classify overcontact binaries in the Kepler samples. Hybrid pipelines that combine multiple methods have also been developed. For instance \citet{kochoska2017gaia} employed polynomial fitting, dimensionality reduction, and DBSCAN clustering to classify Gaia EBs, and \citet{daza2023automated} introduced a Compound Decision Tree framework combining multiple supervised learners using features based on morphology, periods, flux distributions, and structural descriptors. Apart from machine learning approaches, \citep{avvakumova2013eclipsing} introduced supervised schemes based on distributions of observable parameters (e.g., eclipse depths and periods) to classify eclipsing binaries into morphology types.} The most popular among these utilizes light curve folding, polynomial fitting, and dimensionality reduction via locally linear embedding to determine a morphology parameter\citep{Matijevi2012}. While this method is robust and scalable to large datasets, it ignores the inherent variability and consequently the dynamics of the eclipsing binaries.  

We note that nonlinear processes are relevant in modeling stellar variability, and the complex intensity variations 
\citep{kollath1990chaotic,Misra2004ApJa, Misra2006AdSpRa, Harikrishnan2011RAAa}. Recent work has demonstrated that such nonlinear processes could be responsible for the irregular variations observed in close binary stars \citep{George2019,George2020a}. The complex dynamical behavior emerging due to nonlinear processes can be effectively detected using techniques of nonlinear time series analysis \citep{BradleyKantz2015,AH2020}. Among them, the method of recurrence analysis \citep{Marwan2007} is established as an efficient tool in time domain astronomy in recent years. For instance, recurrence analysis of optical light curves was used to detect quasi-periodic oscillations (QPOs) in an active galactic nuclei (AGN) and to distinguish between stochastic, periodic, and chaotic structures underlying the light curves of micro-quasars \citep{Phillipson2020, Sukova2016a}.
 It has also been used to distinguish between the accretion states of the X-ray binary system, GRS~1915+105 \citep{jacob2018recurrence}. In the context of binaries, recurrence analysis has been used to classify binary stars into three morphology classes from their light curves and compared with the second revision of the Kepler EB catalog \citep{George2019,Slawson2011a}.

In this work, we develop an alternate approach to determine the morphology of close binary stars from the dynamics underlying their light curves as quantified using recurrence quantification analysis (RQA). We start by presenting briefly the methods of nonlinear time series analysis and discuss how we apply RQA on observational datasets of close binary stars given in the third revision of the Kepler EB Catalog (KEBC3)~\citep{Matijevi2012,abdul2016kepler}. We study how the RQA measures vary among the different categories of close binary stars and define a Dynamically Derived Morphology (DDM) parameter from them. This novel parameter offers an alternate method for classifying close binary stars from their light curves while accounting for the observed irregular light variations.
\section{Methods}

\label{sect:Obs}
\subsection{Data and Pre-processing}
\label{ssec:data}
For our study, we use data from the Kepler exoplanet exploratory mission \citep{Borucki2010Sci}.
The Kepler mission collects data using a space-based photometric telescope designed to observe a specific patch of the sky over an extended period of four years \mbox{(2009-2013)}. The observations are available in two modes: \textit{long cadence (LC)} and \textit{short cadence (SC)}. For our analysis, we use the LC mode which has a cadence of $\sim$30~minutes. The telescope is in a heliocentric orbit and rotates quarterly to orient its solar panels towards the Sun. As a result, the observations are separated into $\sim$90-day epochs called quarters ($Q0~\mathrm{to}~Q17$). Combining and normalizing these quarters result in a single continuous light curve spanning $\sim$4 years. This data is available at the Mikulski Archive for Space Telescopes (MAST)\footnote{\url{https://mast.stsci.edu/portal/Mashup/Clients/Mast/Portal.html}}. 

\subsubsection*{Target data selection}
Binary stellar systems are one of the object types studied using the capabilities of Kepler mission and the first Kepler Eclipsing Binary Catalog was released in 2010 \citep{Prsa2011a}; a revised catalog was available from 2011 \citep{Slawson2011a} followed by KEBC3\footnote{\label{fnkebc3}\url{https://archive.stsci.edu/kepler/eclipsing_binaries.html}}. This catalog lists 2920 systems classified into detached binaries and close binaries based on a morphology parameter,$c$, defined by \citep{Matijevi2012}. The morphology parameter varies between \mbox{0 to 1}: systems with values between \mbox{0--0.5} are largely detached systems, while the close binaries have values between \mbox{0.5--1}. In this paper,  we choose close binary stars with $c$ between \mbox{0.5--1}. Among them, $c$ between \mbox{0.5--0.7} are largely semi-detached (SD), \mbox{0.7--0.8} are largely overcontact (OC) and \mbox{0.8--1} are largely ellipsoidal (ELV) or have unknown morphology (UN) \citep{Matijevi2012}. Of the 2920 sources in KEBC3, 2907 are unique with 1422 detached binaries, 1311 close binaries and 174 unclassified systems. From the list of 1311, we exclude 8 systems identified as oscillating variables \citep{Borkovits2015mnras} and 66 systems whose light curves are affected by data gaps thus giving us a sample of 1237 close binaries. The 1237 sources consist of 396 sources with $c$ between \mbox{0.5--0.7}, 276 sources with $c$ between \mbox{0.7--0.8} and 565 sources with $c$ between \mbox{0.8--1}. We use the light curves of these sources to study the dynamics of close binary stars.

    

\subsubsection*{Pre-processing of data}
\begin{figure}
    
\begin{subfigure}[b]{0.5\textwidth}
        \centering

    \includegraphics[width=\columnwidth,height=6cm]{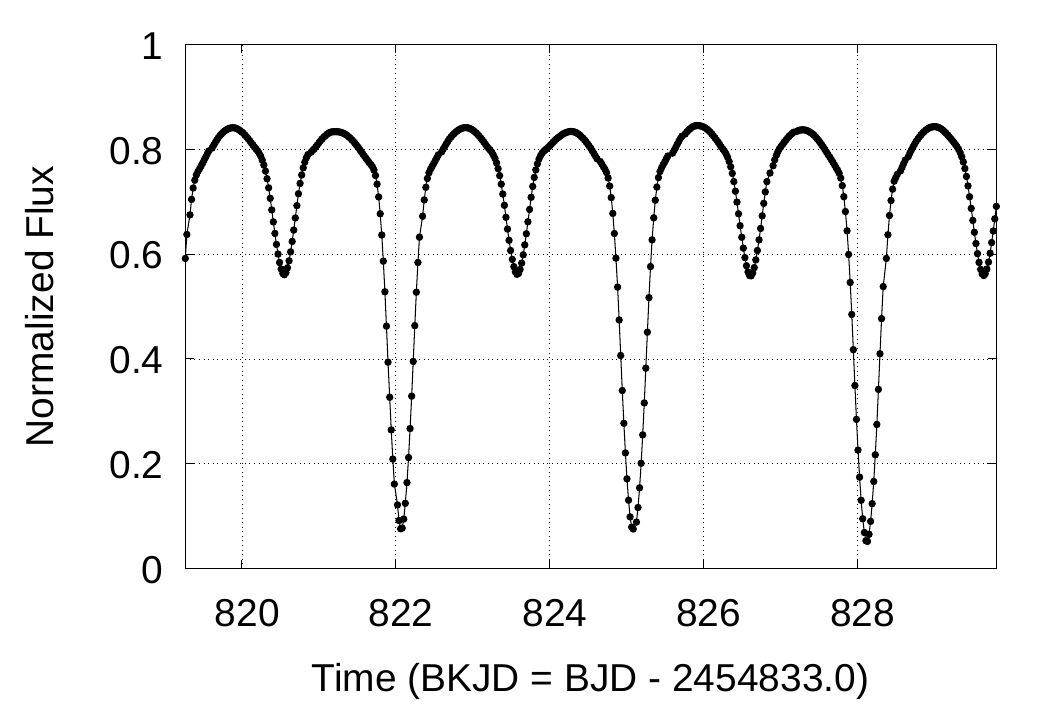}
        \caption{KIC 10014830}
\end{subfigure}
    \vspace{0.5em}

\begin{subfigure}[b]{0.5\textwidth}
        \centering
\includegraphics[width=\columnwidth,height=6cm]{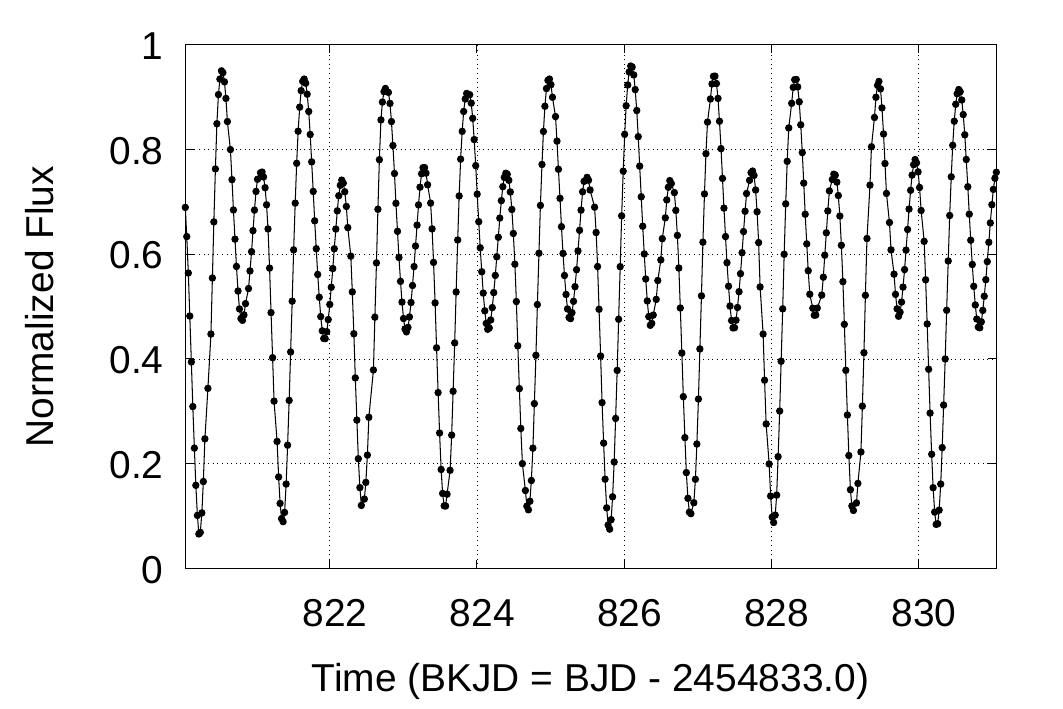}
        \caption{KIC 9164694}
\end{subfigure}  
    \vspace{0.5em}
\begin{subfigure}[b]{0.5\textwidth}
        \centering
   \includegraphics[width=\columnwidth,height=6cm]{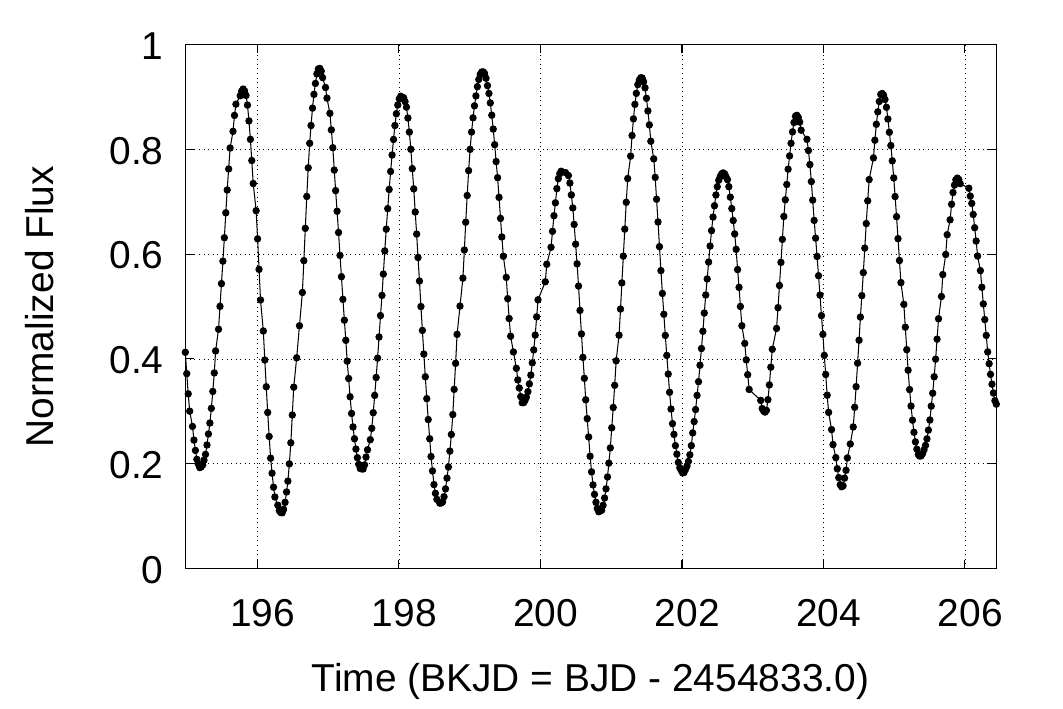}
        \caption{KIC 9544350}
\end{subfigure}

    \caption{Light curve sections ($\sim 10$ days) of three typical binary stars preprocessed using the method described in Section~\ref{ssec:data}. The time axis is given in Barycentric Kepler Julian Date (BKJD) which is the Barycentric Julian Date (BJD) with a zero point of 2454833.0 (BKJD = BJD-2454833).
    \textit{Panel a:} \mbox{KIC 10014830 (morphology parameter, $c=0.61$)}. \textit{Panel b:} \mbox{KIC 9164694 ($c=0.75$)}. \textit{Panel c:} \mbox{KIC 9544350 ($c=0.92$)}.}
    \label{fig:LC}
\end{figure}

For each source selected from KEBC3, we start with the data from the available quarters and filter the data points by setting the quality flag to zero (Qualityflag = 0), effectively excluding any anomalies such as instrumental noise and cosmic ray events. Following this, all quarters are stitched together using the \textit{Lightkurve} package to obtain the light curve with $\sim$60,000 points for each source \citep{Lightkurve2018a, Astropy2022ApJ}.  This light curve is processed through the \mbox{Savitzky-Golay} filter which smoothens the data and improves precision while retaining the signal characteristics\citep{schafer2011savitzky}. We subsequently apply the rolling average method to suppress high-frequency noise caused by instrumental jitter, detector-related brightness variations, and transient cosmic ray events, thereby smoothing short-term fluctuations while preserving the underlying astrophysical features. The data is then rescaled into (0,1) and segments of $\sim$3000 points of the processed data is used as input for further nonlinear analysis. Figure \ref{fig:LC} shows $\sim$10-day segments from typical light curves of systems with varying morphology values, representative of the three classes of close binaries.
\textcolor{red}{The photometric precision of the Kepler mission is expressed in parts per million (ppm), with typical values of the order of a few tens of ppm ($\sim 0.003\%$) on multi-hour time-scales for magnitude 12 stars, which is much smaller than the intrinsic variability in eclipsing binaries stars~\citep{gilliland2011kepler,koch2006kepler,aigrain2015precise}}.



%
\subsection{Recurrences in reconstructed dynamics}
\label{sect:ConstRPnRN}
\begin{figure}
    \centering
    \begin{subfigure}[b]{0.5\textwidth}
        \centering

        \includegraphics[width=\textwidth]{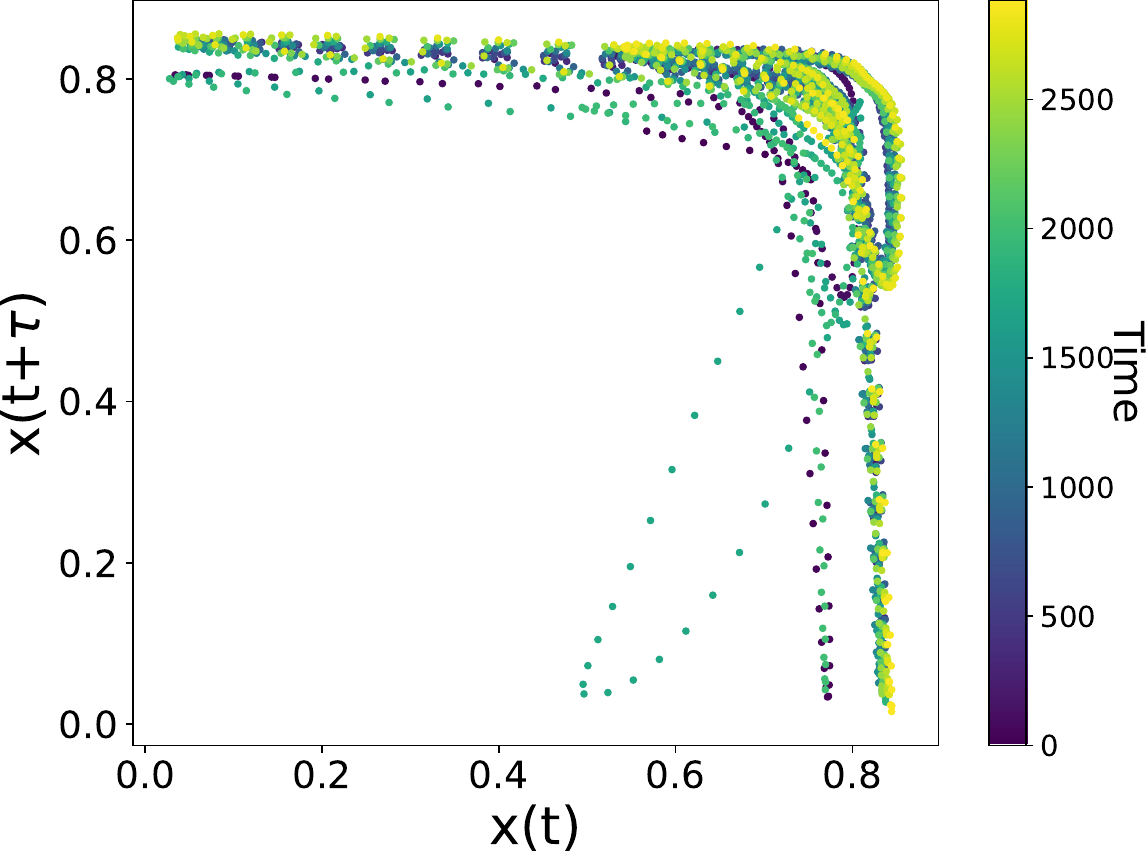}   
        \caption{KIC 10014830}
    \end{subfigure}%
    \hfill
    \begin{subfigure}[b]{0.5\textwidth}
        \centering

        \includegraphics[width=\textwidth]{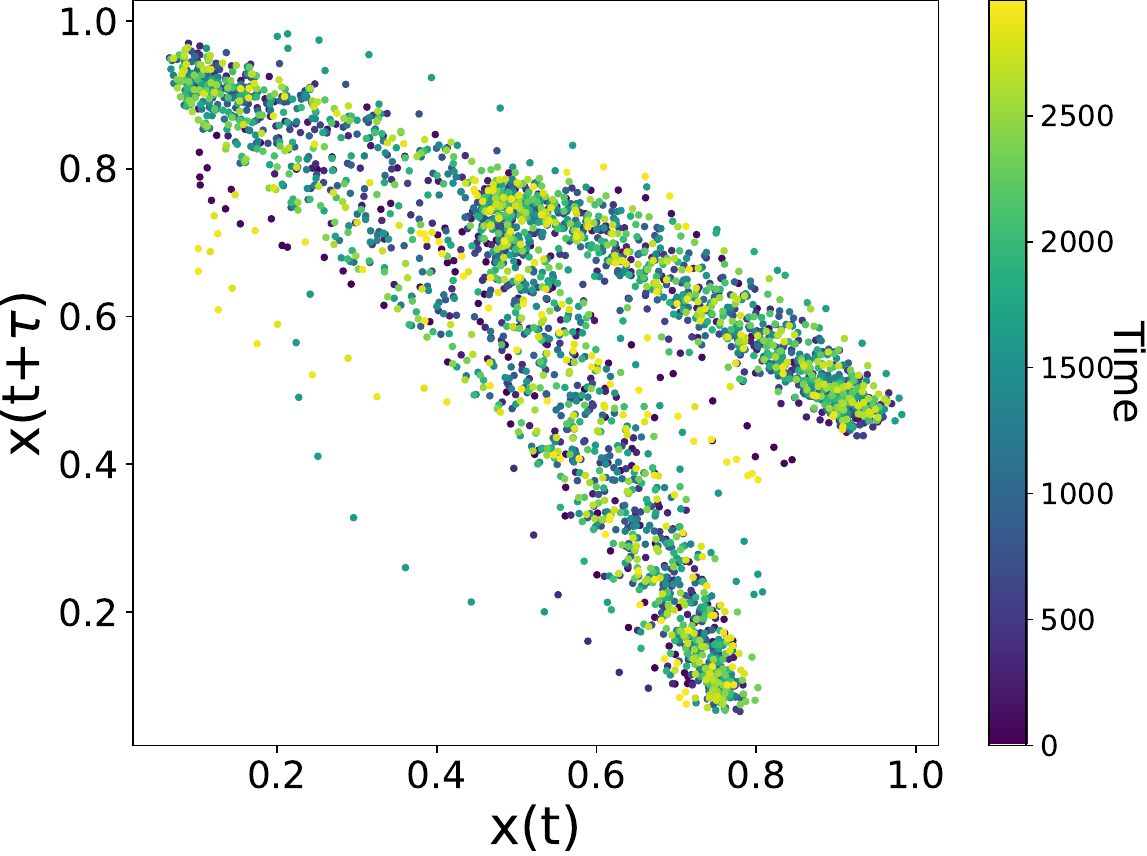}
        \caption{KIC 9164694}
       
    \end{subfigure}

    \vspace{0.5cm} 

    \begin{subfigure}[b]{0.5\textwidth}
        \centering

        \includegraphics[width=\textwidth]{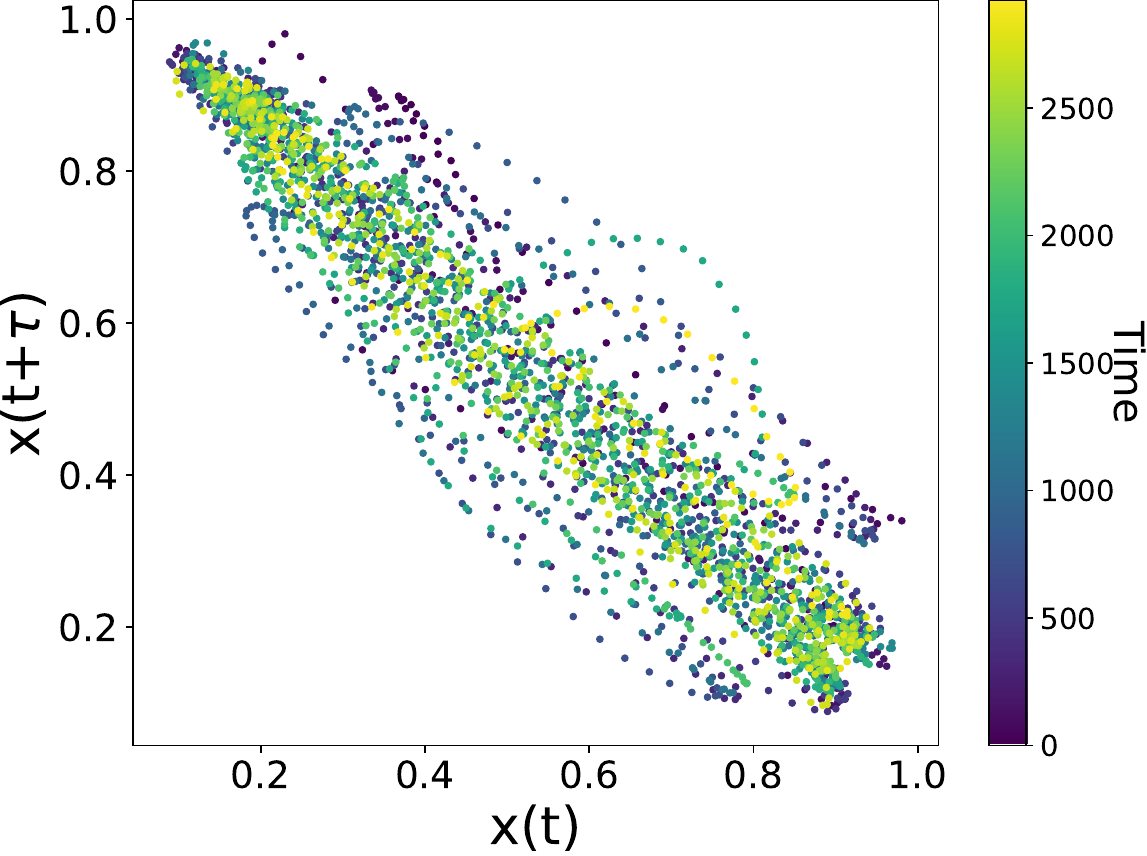}
        \caption{KIC 9544350}
    \end{subfigure}
    \caption{
    Two dimensional projections of reconstructed phase space trajectories for the typical binary stars whose light curves are shown in Figure~\ref{fig:LC}.
    \textit{Panel a:} \mbox{KIC 10014830 (morphology parameter, $c=0.61$)}. \textit{Panel b:} \mbox{KIC 9164694 ($c=0.75$)}. \textit{Panel c:} \mbox{KIC 9544350 ($c=0.92$)}.}
    \label{fig:ReconstPhase}
\end{figure}
We reconstruct the underlying dynamics from the preprocessed data sets using Taken's delay embedding method~\citep{Takens1981}. For this, delay vectors are generated from $N$ scalar data points as $\mathbf{X}_t = [x_t, x_{t+\tau}, \ldots, x_{t+(m-1)\tau}]$ for $t = 1, 2, \ldots, N-m\tau$  to embed in an m-dimensional phase space~\citep{packard1980geometry}. Here, the delay time $\tau$ for embedding is obtained from the first minimum of the autocorrelation function for each data \citep{tan2023selecting}, while the embedding dimension $m$, is estimated using the method of False Nearest Neighbours (FNN) \citep{kennel1992determining}. In our analysis, we set $m=4$, since that is the value obtained for the majority of the data sets. 

The dynamics thus reconstructed in the m-dimensional phase space can be projected in 2-dimensions $(x(t)$ vs $x(t + \tau ))$, for visualization, as shown in Figure~\ref{fig:ReconstPhase} for the stars included in Figure~\ref{fig:LC}. For any bounded system, the embedded trajectory visits the same region of the phase space several times and the resulting recurrence pattern is characteristic of the underlying dynamics of the system. 

The Recurrence Plot (RP)~\citep{Eckmann1995} constructed from the delay vectors captures these recurrence patterns~\citep{Marwan2007}. For this, we compute the recurrence matrix ($R$) that encodes the recurrence relationships between delay vectors, with elements of binary values (0 or 1) as follows:
\begin{equation}\label{rij01}
    R_{ij} = \Theta(\varepsilon - \| \mathbf{X}_i - \mathbf{X}_j \|)
\end{equation}
where $\Theta$ is the Heaviside step function, $\| \cdot \|$ denotes the Euclidean distance between the two points $X_i$ and $X_j$and $\epsilon$ defines the threshold distance for the points to be considered "recurring". For the chosen embedding dimension $m=4$, we take  $\epsilon=0.16$, to ensure comparability across all datasets.. 





\begin{figure}
    \centering
    \begin{subfigure}[b]{0.4\textwidth}
        \includegraphics[width=\textwidth]{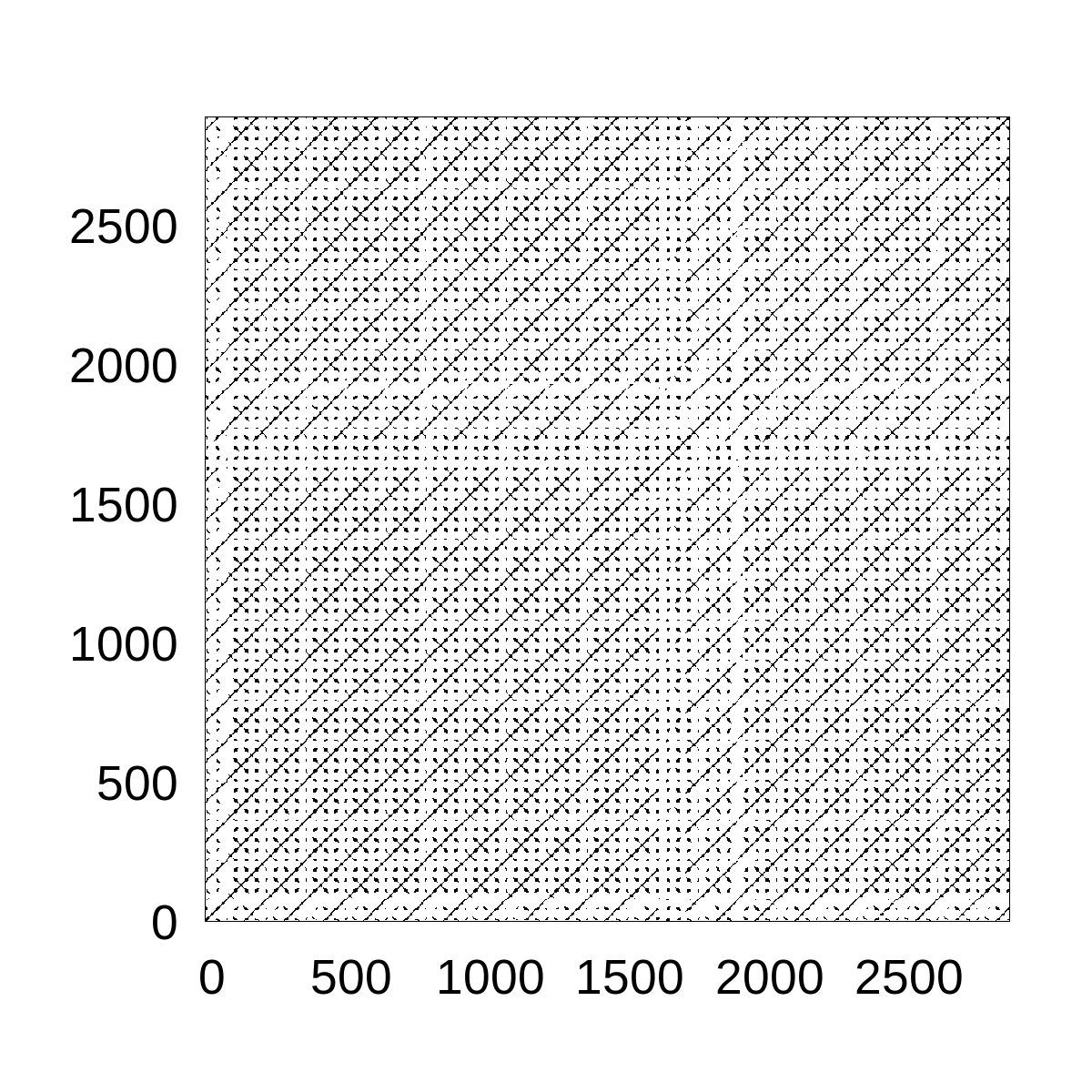}
        \caption{KIC 10014830}
    \end{subfigure}%
    \hfill
    \begin{subfigure}[b]{0.4\textwidth}
        \centering

    \includegraphics[width=\textwidth]{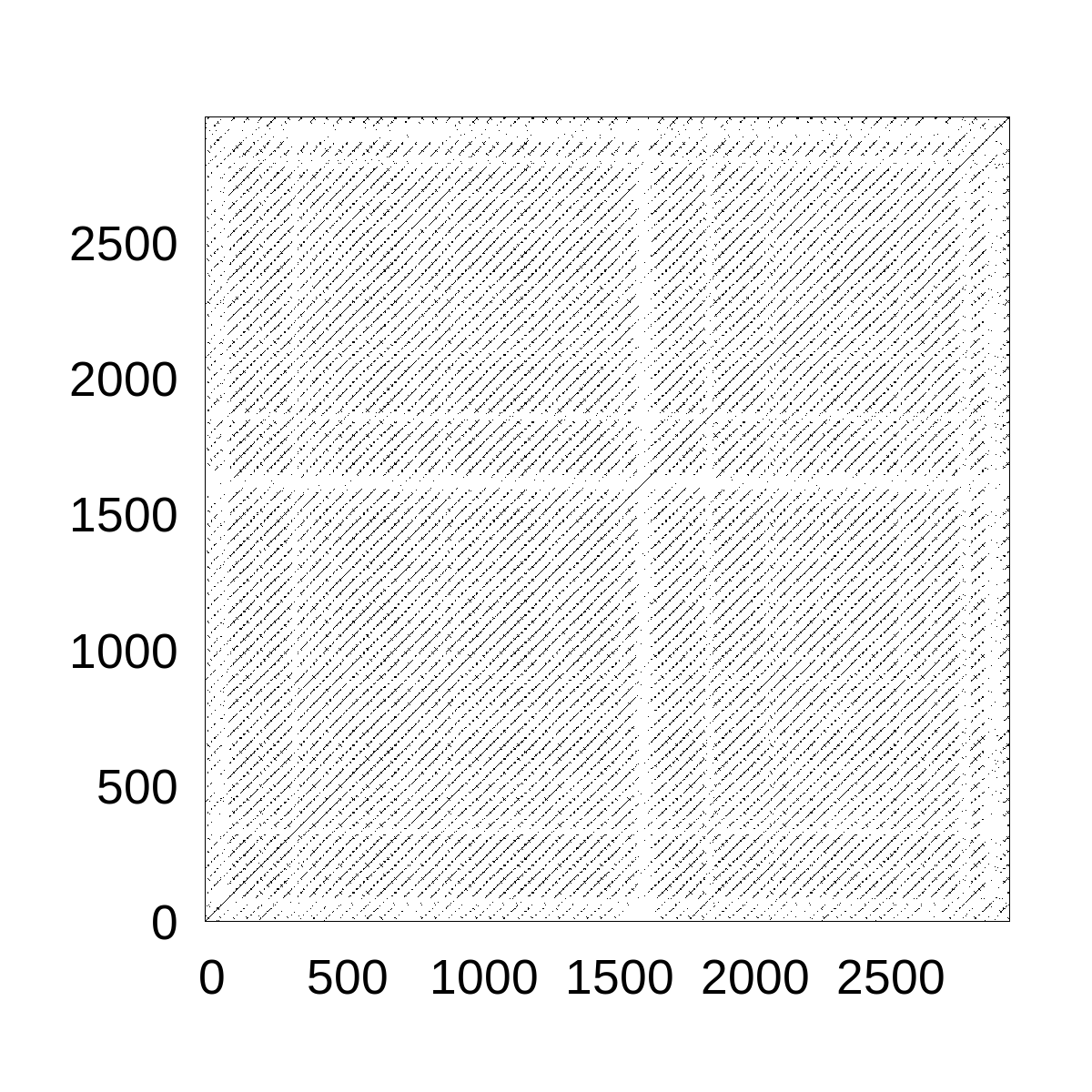}
        \caption{KIC 9164694}
    \end{subfigure}

    \vspace{0.5cm} 

    \begin{subfigure}[b]{0.4\textwidth}
        \centering

        \includegraphics[width=\textwidth]{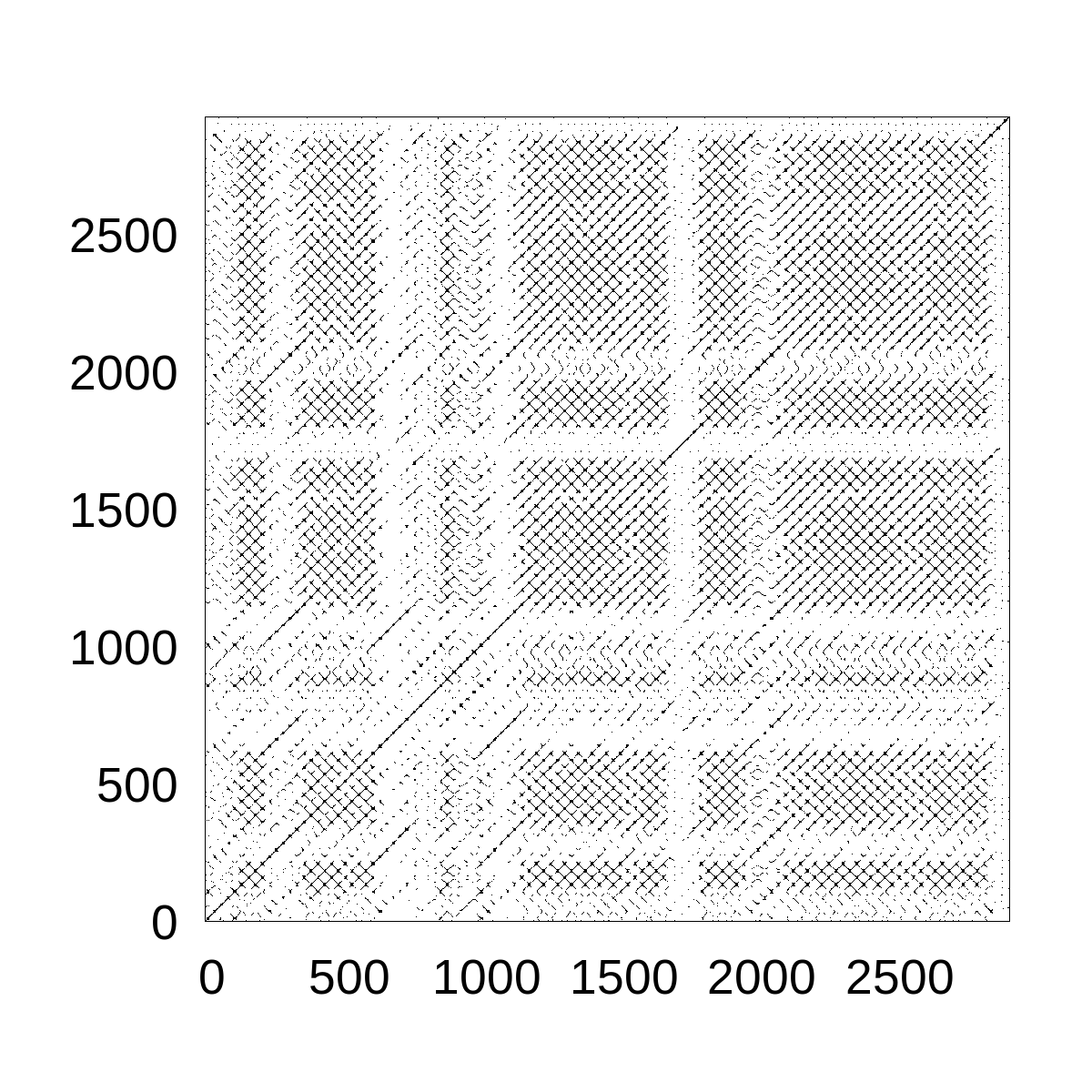}
        \caption{KIC 9544350}
    \end{subfigure}
    \caption{Recurrence plots from the reconstructed trajectories of the three different types of close binary stars shown in Figure ~\ref{fig:ReconstPhase}. \textit{Panel a:} \mbox{KIC 10014830 ($c=0.61$)}. \textit{Panel b:} \mbox{KIC 9164694 ($c=0.75$)}. \textit{Panel c:} \mbox{KIC 9544350 ($c=0.92$)}.}
    \label{fig:RPs}
\end{figure}

The Recurrence plot is a visual representation of the matrix $R$, with dots marked corresponding to nonzero elements in $R$. Thus, the dots in the plot indicate that the state of the system at a time $i$ has recurred at a later time $j$ 
In general the RPs show patterns formed by isolated points, diagonal, vertical and horizontal lines, each of which provides an insight into the system's underlying dynamics. The recurrence plots (RPs) derived from the embedded trajectories of the same systems in Figure~\ref{fig:LC}, are shown in Figure~\ref{fig:RPs}. It is clear that they exhibit varying patterns that reflect differences in their morphology. 


To quantify these patterns, we compute recurrence measures such as Recurrence Rate (RR), Determinism ($DET$), Laminarity (LAM) and Entropy ($ENT$)~\citep{Marwan2007}. 
$RR$ gives an estimate of the density of points in RP or the fraction of recurrent points as
 \begin{equation}\label{rr01}
    \text{RR} = \frac{1}{N^2} \sum_{i,j=1}^{N} R_{ij}
\end{equation}
Here $N$ is the total number of points on the reconstructed phase space trajectory.
 
$DET$ quantifies the deterministic aspect of the system's dynamics by measuring the proportion of recurrent points that form diagonal lines in the recurrence plot. 
It is calculated as
\begin{equation}\label{det01}
    \text{DET} = \frac{\sum_{l=l_{\text{min}}}^{N} l \cdot P(l)}{\sum_{l=1}^{N} l \cdot P(l)}   
\end{equation}
Here $l$ is the length of a diagonal lines in the recurrence plot and $P(l)$ is the corresponding probability distribution. The parameter $l_{\text{min}}$ denotes the minimum length of the diagonal lines that are considered for the calculation of $DET$. In this analysis, we used the standard default $l_{\text{min}}=2$ \citep{BABAEI2014112}, to exclude short, noise-induced diagonal lines \citep{Marwan2007}. 
It is known that high $DET$ values indicate periodicity or regularity, implying deterministic behavior, while low $DET$ values suggest more irregular and stochastic dynamics. 

LAM quantifies the presence of vertical recurrence lines, reflecting persistent states within the time series and is calculated using
\begin{equation}\label{lam01}
    \text{LAM} = \frac{\sum_{v=v_{\text{min}}}^{N} v \cdot P(v)}{\sum_{v=1}^{N} v \cdot P(v)}   
\end{equation}
Here $v$ represents the length of a vertical line in the recurrence plot and $P(v)$ is the probability distribution of the length $v$ of vertical lines. The parameter $v_{\text{min}}$  defines the minimum length of the vertical lines considered for LAM. In this study, $v_{\text{min}}$ is set to the default value 2.
A high LAM value suggests that the system tends to remain in similar states for several consecutive time steps, exhibiting short-term stability before transitioning. Conversely, a low LAM value indicates that the system frequently switches between different states, without lingering in any particular state for extended periods.

$ENT$ measures the degree of uncertainty in the length distribution of the diagonal lines and is defined as:  

\begin{equation}
    \text{ENT} = -\sum_{l = l_{\min}}^{N} p(l) \log p(l)  \label{ent01}
\end{equation}
Here,  $p(l)$ is the normalized probability of a diagonal line of length $l$, given by:
\begin{equation}
    p(l) = \frac{P(l)}{\sum_{l = l_{\min}}^{N} P(l)}  \tag*{} \label{ent_pl01}
\end{equation}
A higher $ENT$ value corresponds to a more complex, unpredictable, or chaotic system, while a lower $ENT$ value indicates a more regular or stochastic system. 
These measures provide a robust framework for classifying systems based on the recurrence pattern of their underlying dynamics. 

\noindent

\noindent


\subsection{Dynamically Derived Morphology (DDM)}
\label{sec:ddm}
Since the differences in the underlying dynamics of each type of the binary stars reflect in the variations in their light curves, the recurrence plot measures can capture distinct aspects of their dynamics, and thus help to classify the binary stars. Instead of considering each recurrence measure separately for classification, \textcolor{red}{ we mainly rely on values of $DET$ and $ENT$ in arriving at a classification index.  $DET$ measures the proportion of recurrent points that form diagonal lines and so high values of $DET$ indicate predictable or more regular dynamics. $ENT$ is the probability distribution of the diagonal line lengths and high $ENT$ indicate a large variety of diagonal line lengths and so high complexity in dynamics.  For deterministic nonlinear systems with highly irregular dynamics,  both $DET$ and $ENT$ will be high.  For periodic signals, $DET$ is high but $ENT$ is small while for uncorrelated noise both are small. Therefore, the combination of both $DET$ and $ENT$ can provide valuable information about the different types of dynamics underlying the light curves of binary stars and hence can lead to classification.} We introduce a new parameter, Dynamically Derived Morphology (DDM) parameter, based on the observed relationship between the $DET$ and $ENT$ measures among the close binaries. To derive the DDM parameter, we fit the following functional form to the plot of all binaries on the $DET$-$ENT$ plane, where $DET$ ($f(x)$) is expressed in terms of $ENT$ ($x$).
\begin{equation}
    f(x)=\frac{1}{1+(\frac{x+k}{x_0})^{-\alpha}}
    \label{eq:fit}
\end{equation}
Here $\alpha$, $k$ and $x_0$ are the parameters of the fit. To reduce the effect of outliers, a robust regression with the Huber loss function is used \citep{huber1992robust}. Subsequently, for each binary, the closest point on ($ENT$, $DET$) curve is identified as ($x^*$,$f(x^*)$). The length on the curve from $0$ to $f(x^*)$ is defined as the DDM parameter for that star.


\section{Results \& Discussion}
\label{sect:Res}

From the reconstructed trajectories of close binary stars, we obtain their recurrence plots and compute the measures $RR$, $DET$, $LAM$ and $ENT$. The statistical analysis of these measures, summarized in Table~\ref{tab:RPmeasure_summary}, is performed after first grouping the binaries into classes based on the existing morphology parameter. Within each class, we report the mean, KDE mode, and standard deviation of each measure.
We analyze how the distributions of these recurrence measures vary between the binary classes. The variations in the mean, and KDE mode values of these metrics highlight distinctions in the underlying dynamical regimes of each class.  \textcolor{red}{As mentioned in Section \ref{sec:ddm}, we focus on $DET$ and $ENT$ as the most relevant measures capturing the predictability and diversity of recurrent phase space structures. We conduct an extensive surrogate analysis on each dataset, comparing the $DET$ and $ENT$ values for the original light curve to $100$ Iterated Amplitude Adjusted Fourier Transform (IAAFT) surrogates, to test whether the observed values are beyond those of a linear stochastic process with the same power spectrum and amplitude distribution \citep{theiler1992testing,schreiber1996improved,Donges2015pyunicorn}. We find for 94\% of stars the $ENT$ values significantly ($p<.05$) differ from that of their surrogates , whereas for 96\% of stars $DET$ values differ significantly ($p<.05$).}

\begin{figure}
    \centering
   
        \includegraphics[width=.5\textwidth]{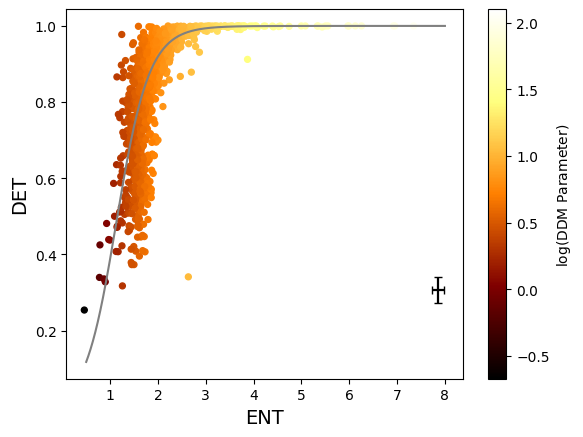}
       
    \caption{\textcolor{red}{$ENT$-$DET$ plane for all the  binary stars mentioned in subsection~\ref{ssec:data} with the best-fit curve in grey. The points are colour coded with the values of the DDM parameter. The typical errorbars on the $DET$ and $ENT$ values are shown in the lower right corner of the plot.}}
    \label{fig:dynparam}
\end{figure}

The values of these two measures for all the close binary stars used in the study are then plotted in the $ENT-DET$ plane, where they are found to lie along a distinct curve. \textcolor{red}{We see that a number of stars show different $DET$ values with the same value of $ENT$, and few others show different $ENT$ values with the same $DET$. Hence, we fit} these values using the function given in Equation \ref{eq:fit}, \textcolor{red}{and obtain the best fit parameters as \textcolor{red}{$x_0=9.99$, $k=8.85$ and $\alpha=29.91$}.} The DDM parameter is then computed for each star from its closest point on the curve. \textcolor{red}{The uncertainties in both $DET$ and $ENT$ are estimated via bootstrap resampling of the diagonal line length distributions (1000 resamples per dataset) following \citet{schinkel2009confidence}.}  The plot of all the binaries on the $ENT$-$DET$ plane along with the best fit curve\textcolor{red}{, and typical errorbars} is shown in Figure~\ref{fig:dynparam}. The points are color coded as per the values of their DDM parameters.
 
 We compare the DDM parameter values with the existing morphology parameter, \textcolor{red}{and observe a small inverse correlation (Spearman $\rho=-0.21$, p-value<$10^{-14})$. Accordingly, stars with higher DDM values appear somewhat more common among the semi-detached systems (lower morphology parameter), while ellipsoidal and over-contact stars (higher morphology parameter) show correspondingly lower DDM values\citep{Matijevi2012}. However, this trend is weak and is best viewed as a descriptive pattern rather than a strong physical distinction. The low value of correlation} indicates that the DDM parameter contains information complementary to the morphology parameter. This is expected, since the morphology parameter relies on the characteristic of the folded light curve, which averages out the variations between eclipses that arise from underlying the nonlinear dynamics.

\begin{table}
\centering
\begin{tabular}{llcccc}
\hline
\textbf{Morphology} & \textbf{Measure} &  \textbf{Mean} & \textbf{KDE mode} & \textbf{SD} \\
\textbf{parameter ($c$)} &  &$(\mu)$  &  & $(\sigma)$ \\
\hline			

0.5-0.7 & $RR$  & 0.20 & 0.11 & 0.20\\ 
                  & $DET$  & 0.90 & 0.89 & 0.09\\ 
                  & $LAM$  & 0.95 & 0.99 & 0.06 \\
                  & $ENT$  & 2.21 & 1.90 & 0.99\\
0.7-0.8 & $RR$   & 0.23 & 0.12 & 0.20 \\
                 & $DET$  & 0.76 & 0.93 & 0.16 \\
                 & $LAM$  & 0.80 & 0.95 & 0.19 \\
                 & $ENT$  & 1.85 & 1.65 & 0.65 \\
0.8-1.0 & $RR$   & 0.24 & 0.11 & 0.22 \\
                 & $DET$  & 0.79 & 0.96 & 0.17 \\
                 & $LAM$  & 0.83 & 0.97 & 0.19\\ 
                 & $ENT$  & 1.96 & 1.69 & 0.87 \\
\hline
\end{tabular}
\caption{Statistics of Recurrence Plot measures for close binary stars in the Kepler Eclipsing Binary Catalog (Revision 3).} 
\label{tab:RPmeasure_summary}
\end{table}

\noindent
   
       

\section{Conclusion}
\label{concl}

The light curves of eclipsing binary stars have a well defined periodic variability. However, in close binary stars these variations may be superimposed with changes in the intensity due to reasons such as tidal disruptions, long lived spots, turbulent mass exchanges or intrinsic stellar dynamics~\citep{fabry2022modeling, fabry2023modeling, cherepashchuk2022close}. Each of these can result in nonlinear fluctuations in the light intensity measured from these stars. In this study, the differences in the nature of these intensity fluctuations are captured by reconstructing the dynamics and using the recurrence measures such as RR, $DET$, LM and $ENT$. A new morphology parameter is introduced that contains information on the underlying dynamics as captured through its recurrence patterns. This DDM parameter offers an alternate classification scheme for close binary stars. In addition, this method is based on the analysis of $\sim$3000 points of the light curve and is hence computationally efficient for large data sets. Further,  analysis with light curves with  larger number of points or averaging multiple segments of $\sim$3000 points from various epochs, do not appreciably change our results. 

With highly sensitive light curves available from telescopes such as Kepler and TESS, our understanding of the variability of close binary stars has improved considerably. Recent studies have shown non trivial patterns in the variations of eclipse times as well as in the O' Connell effect (uneven amplitudes for out of eclipse maxima)~\citep{tran2013anticorrelated, wang2015flip, knote2022characteristics}.  These indicate that the dynamics beyond periodic fluctuations in eclipsing binaries is significant, and quantifying these may be crucial in increasing our understanding of their underlying mechanisms.

\textcolor{red}{We note the physical interactions in close binary stars, like dynamical tides, episodic or unstable mass and energy transfer, variations in accretion structures or hot spots, and thermal–relaxation oscillations, can naturally lead to rich nonlinear dynamics~\citep{fabry2022modeling,fabry2023modeling,qian2001orbital}. These processes often produce cycle-to-cycle variability or quasi-periodic behaviour that does not repeat coherently from orbit to orbit~\citep{qian2003overcontact,sriram2017study,li2024detection}. As a result, their signatures may be visible in the light curve but are largely suppressed or entirely lost in the folding procedure used to derive morphology. In this context, the weak relationship we observe between DDM and morphology suggests that the DDM parameter is sensitive to dynamical behaviour that is not captured by the folded light curve and may therefore reflect the influence of these nonlinear physical processes.}

As reported earlier, the metrics derived from nonlinear time series analysis can be related to the morphology and other related astrophysical parameters for close binaries \citep{George2019,George2020a}.  The present study goes a step further by defining a new parameter which can identify close binary stars exhibiting similar nonlinear dynamics. Hence, our method, based on nonlinear analysis of light curves, introduces a new approach towards classification of close binary stars.

Astronomical Sky Surveys result in large data sets that contain light curves of millions of sources, a considerable number of them, being variables. These sources are usually classified using methods based on spectral and time domain analysis. We show that we can classify close binary stars based solely on the  parameters obtained from the nonlinear time series analysis of their light curves. Since this method is based on photometric variability and is computationally optimal, we plan to improve it further for better accuracy and compatibility with ongoing and upcoming astronomy missions.

\section*{Acknowledgements}
This paper includes data collected by the Kepler mission and obtained from the MAST data archive at the Space Telescope Science Institute (STScI). Funding for the Kepler mission is provided by the NASA Science Mission Directorate. STScI is operated by the Association of Universities for Research in Astronomy, Inc., under NASA contract NAS 5–26555.
DP thanks ISRO-INDIA for support through the ISRO-RESPOND programme. DP, GA and SVG thank IUCAA for supporting collaborations and meetings. AK thanks IUCAA for support through a Post-Doctoral Fellowship, funded by the University Grants Commission (UGC).  

 

\section*{Data and Software Availability}

All the data sets used in the study are from the Kepler exoplanetary search mission (\url{https://archive.stsci.edu/missions-and-data/kepler}) and these data are available at the Mikulski Archive for Space Telescopes (\url{https://mast.stsci.edu/portal/Mashup/Clients/Mast/Portal.html}). The revised Kepler eclipsing binary catalog is available at \url{https://keplerebs.villanova.edu/}. 

The analysis of preprocessed light curves (Section~\ref{ssec:data}) are done using programs written in Python, including scientific libraries such as NumPy and SciPy, Astropy, Lightkurve, and Matplotlib and modules from Kaggle\footnote{\url{https://www.kaggle.com/}}.
For recurrence analysis using RQA we used tools from PyUnicorn\footnote{\url{https://www.pik-potsdam.de/members/donges/software-2/software}} developed by Donges et al.~\citep{Donges2015pyunicorn}. 





\begin{thebibliography}{}
\makeatletter
\relax
\def\mn@urlcharsother{\let\do\@makeother \do\$\do\&\do\#\do\^\do\_\do\%\do\~}
\def\mn@doi{\begingroup\mn@urlcharsother \@ifnextchar [ {\mn@doi@} {\mn@doi@[]}}
\def\mn@doi@[#1]#2{\def\@tempa{#1}\ifx\@tempa\@empty \href {http://dx.doi.org/#2} {doi:#2}\else \href {http://dx.doi.org/#2} {#1}\fi \endgroup}
\def\mn@eprint#1#2{\mn@eprint@#1:#2::\@nil}
\def\mn@eprint@arXiv#1{\href {http://arxiv.org/abs/#1} {{\tt arXiv:#1}}}
\def\mn@eprint@dblp#1{\href {http://dblp.uni-trier.de/rec/bibtex/#1.xml} {dblp:#1}}
\def\mn@eprint@#1:#2:#3:#4\@nil{\def\@tempa {#1}\def\@tempb {#2}\def\@tempc {#3}\ifx \@tempc \@empty \let \@tempc \@tempb \let \@tempb \@tempa \fi \ifx \@tempb \@empty \def\@tempb {arXiv}\fi \@ifundefined {mn@eprint@\@tempb}{\@tempb:\@tempc}{\expandafter \expandafter \csname mn@eprint@\@tempb\endcsname \expandafter{\@tempc}}}

\bibitem[\protect\citeauthoryear{Abdul-Masih et~al.,}{Abdul-Masih et~al.}{2016}]{abdul2016kepler}
Abdul-Masih M.,  et~al., 2016, The Astronomical Journal, 151, 101

\bibitem[\protect\citeauthoryear{Aigrain, Hodgkin, Irwin, Lewis  \& Roberts}{Aigrain et~al.}{2015}]{aigrain2015precise}
\textcolor{red}{Aigrain S.,  Hodgkin S.~T.,  Irwin M.~J.,  Lewis J.~R.,   Roberts S.~J.,  2015, Monthly notices of the royal astronomical society, 447, 2880}

\bibitem[\protect\citeauthoryear{Ambika \& Harikrishnan}{Ambika \& Harikrishnan}{2020}]{AH2020}
Ambika G.,  Harikrishnan K.,  2020, in Mukhopadhyay A.,  Sen S.,  Basu D.~N.,   Mondal S.,  eds, , Dynamics and Control of Energy Systems.
Springer, Singapore, pp 9--27

\bibitem[\protect\citeauthoryear{{Astropy Collaboration}}{{Astropy Collaboration}}{2022}]{Astropy2022ApJ}
{Astropy Collaboration} 2022, \mn@doi [\apj] {10.3847/1538-4357/ac7c74}, \href {https://ui.adsabs.harvard.edu/abs/2022ApJ...935..167A} {935, 167}

\bibitem[\protect\citeauthoryear{Avvakumova, Malkov  \& Kniazev}{Avvakumova et~al.}{2013}]{avvakumova2013eclipsing}
\textcolor{red}{Avvakumova E.,  Malkov O.~Y.,   Kniazev A.~Y.,  2013, Astronomische Nachrichten, 334, 860}

\bibitem[\protect\citeauthoryear{Babaei, Zarghami, Sedighikamal, Sotudeh-Gharebagh  \& Mostoufi}{Babaei et~al.}{2014}]{BABAEI2014112}
Babaei B.,  Zarghami R.,  Sedighikamal H.,  Sotudeh-Gharebagh R.,   Mostoufi N.,  2014, \mn@doi [Physica A: Statistical Mechanics and its Applications] {https://doi.org/10.1016/j.physa.2013.10.016}, 395, 112

\bibitem[\protect\citeauthoryear{Borkovits, Hajdu, Sztakovics, Rappaport, Levine, Bíró  \& Klagyivik}{Borkovits et~al.}{2015}]{Borkovits2015mnras}
Borkovits T.,  Hajdu T.,  Sztakovics J.,  Rappaport S.,  Levine A.,  Bíró I.~B.,   Klagyivik P.,  2015, \mn@doi [Monthly Notices of the Royal Astronomical Society] {10.1093/mnras/stv2530}, 455, 4136

\bibitem[\protect\citeauthoryear{{Borucki} et~al.,}{{Borucki} et~al.}{2010}]{Borucki2010Sci}
{Borucki} W.~J.,  et~al., 2010, \mn@doi [Science] {10.1126/science.1185402}, \href {https://ui.adsabs.harvard.edu/abs/2010Sci...327..977B} {327, 977}

\bibitem[\protect\citeauthoryear{Bradley \& Kantz}{Bradley \& Kantz}{2015}]{BradleyKantz2015}
Bradley E.,  Kantz H.,  2015, \mn@doi [Chaos: An Interdisciplinary Journal of Nonlinear Science] {10.1063/1.4917289}, 25, 097610

\bibitem[\protect\citeauthoryear{Cherepashchuk}{Cherepashchuk}{2022}]{cherepashchuk2022close}
Cherepashchuk A.,  2022, Astronomy Reports, 66, S5

\bibitem[\protect\citeauthoryear{{\v{C}}okina, Maslej-Kre{\v{s}}{\v{n}}{\'a}kov{\'a}, Butka  \& Parimucha}{{\v{C}}okina et~al.}{2021a}]{vcokina2021automatic}
\textcolor{red}{{\v{C}}okina M.,  Maslej-Kre{\v{s}}{\v{n}}{\'a}kov{\'a} V.,  Butka P.,   Parimucha {\v{S}}.,  2021a, Astronomy and Computing, 36, 100488}

\bibitem[\protect\citeauthoryear{{\v{C}}okina, Fedurco  \& Parimucha}{{\v{C}}okina et~al.}{2021b}]{vcokina2021elisa}
\textcolor{red}{{\v{C}}okina M.,  Fedurco M.,   Parimucha {\v{S}}.,  2021b, Astronomy \& Astrophysics, 652, A156}

\bibitem[\protect\citeauthoryear{Daza-Perilla, Gramajo, Lares, Palma, Ferreira~Lopes, Minniti  \& Clari{\'a}}{Daza-Perilla et~al.}{2023}]{daza2023automated}
\textcolor{red}{Daza-Perilla I.~V.,  Gramajo L.~V.,  Lares M.,  Palma T.,  Ferreira~Lopes C.,  Minniti D.,   Clari{\'a} J.,  2023, Monthly Notices of the Royal Astronomical Society, 520, 828}

\bibitem[\protect\citeauthoryear{Donges et~al.,}{Donges et~al.}{2015}]{Donges2015pyunicorn}
Donges J.~F.,  et~al., 2015, \mn@doi [Chaos: An Interdisciplinary Journal of Nonlinear Science] {10.1063/1.4934554}, 25, 113101

\bibitem[\protect\citeauthoryear{Eckmann, Kamphorst, Ruelle  et~al.}{Eckmann et~al.}{1995}]{Eckmann1995}
Eckmann J.-P.,  Kamphorst S.~O.,  Ruelle D.,   et~al., 1995, World Scientific Series on Nonlinear Science Series A, 16, 441

\bibitem[\protect\citeauthoryear{Fabry, Marchant  \& Sana}{Fabry et~al.}{2022}]{fabry2022modeling}
Fabry M.,  Marchant P.,   Sana H.,  2022, Astronomy \& Astrophysics, 661, A123

\bibitem[\protect\citeauthoryear{Fabry, Marchant, Langer  \& Sana}{Fabry et~al.}{2023}]{fabry2023modeling}
Fabry M.,  Marchant P.,  Langer N.,   Sana H.,  2023, Astronomy \& Astrophysics, 672, A175

\bibitem[\protect\citeauthoryear{George, Misra  \& Ambika}{George et~al.}{2019}]{George2019}
George S.~V.,  Misra R.,   Ambika G.,  2019, \mn@doi [Chaos: An Interdisciplinary Journal of Nonlinear Science] {10.1063/1.5120739}, 29, 113112

\bibitem[\protect\citeauthoryear{George, Misra  \& Ambika}{George et~al.}{2020}]{George2020a}
George S.~V.,  Misra R.,   Ambika G.,  2020, \mn@doi [Communications in Nonlinear Science and Numerical Simulation] {https://doi.org/10.1016/j.cnsns.2019.104988}, 80, 104988

\bibitem[\protect\citeauthoryear{Gilliland et~al.,}{Gilliland et~al.}{2011}]{gilliland2011kepler}
\textcolor{red}{Gilliland R.~L.,  et~al., 2011, The Astrophysical Journal Supplement Series, 197, 6}

\bibitem[\protect\citeauthoryear{{Harikrishnan}, {Misra}  \& {Ambika}}{{Harikrishnan} et~al.}{2011}]{Harikrishnan2011RAAa}
{Harikrishnan} K.~P.,  {Misra} R.,   {Ambika} G.,  2011, \mn@doi [Research in Astronomy and Astrophysics] {10.1088/1674-4527/11/1/004}, \href {https://ui.adsabs.harvard.edu/abs/2011RAA....11...71H} {11, 71}

\bibitem[\protect\citeauthoryear{Huber}{Huber}{1992}]{huber1992robust}
Huber P.~J.,  1992, in , Breakthroughs in statistics: Methodology and distribution.
Springer, pp 492--518

\bibitem[\protect\citeauthoryear{Jacob, Harikrishnan, Misra  \& Ambika}{Jacob et~al.}{2018}]{jacob2018recurrence}
Jacob R.,  Harikrishnan K.,  Misra R.,   Ambika G.,  2018, Communications in Nonlinear Science and Numerical Simulation, 54, 84

\bibitem[\protect\citeauthoryear{Kennel, Brown  \& Abarbanel}{Kennel et~al.}{1992}]{kennel1992determining}
Kennel M.~B.,  Brown R.,   Abarbanel H.~D.,  1992, Physical review A, 45, 3403

\bibitem[\protect\citeauthoryear{Knote, Caballero-Nieves, Gokhale, Johnston  \& Perlman}{Knote et~al.}{2022}]{knote2022characteristics}
Knote M.~F.,  Caballero-Nieves S.~M.,  Gokhale V.,  Johnston K.~B.,   Perlman E.~S.,  2022, The Astrophysical Journal Supplement Series, 262, 10

\bibitem[\protect\citeauthoryear{Koch et~al.,}{Koch et~al.}{2006}]{koch2006kepler}
\textcolor{red}{Koch D.,  et~al., 2006, Proceedings of the International Astronomical Union, 2, 236}

\bibitem[\protect\citeauthoryear{Kochoska, Mowlavi, Pr{\v{s}}a, Lecoeur-Ta{\"\i}bi, Holl, Rimoldini, S{\"u}veges  \& Eyer}{Kochoska et~al.}{2017}]{kochoska2017gaia}
Kochoska A.,  Mowlavi N.,  Pr{\v{s}}a A.,  Lecoeur-Ta{\"\i}bi I.,  Holl B.,  Rimoldini L.,  S{\"u}veges M.,   Eyer L.,  2017, Astronomy \& Astrophysics, 602, A110

\bibitem[\protect\citeauthoryear{Koll{\'a}th}{Koll{\'a}th}{1990}]{kollath1990chaotic}
Koll{\'a}th Z.,  1990, Monthly Notices of the Royal Astronomical Society, Vol. 247, NO. 3/DEC1, P. 377, 1990, 247, 377

\bibitem[\protect\citeauthoryear{Li et~al.,}{Li et~al.}{2024}]{li2024detection}
\textcolor{red}{Li M.-Y.,  et~al., 2024, The Astrophysical Journal, 962, 44}

\bibitem[\protect\citeauthoryear{{Lightkurve Collaboration} et~al.,}{{Lightkurve Collaboration} et~al.}{2018}]{Lightkurve2018a}
{Lightkurve Collaboration} et~al., 2018, {Lightkurve: Kepler and TESS time series analysis in Python}, Astrophysics Source Code Library (\mn@eprint {ascl} {1812.013})

\bibitem[\protect\citeauthoryear{Marwan, {Carmen Romano}, Thiel  \& Kurths}{Marwan et~al.}{2007}]{Marwan2007}
Marwan N.,  {Carmen Romano} M.,  Thiel M.,   Kurths J.,  2007, \mn@doi [Physics Reports] {https://doi.org/10.1016/j.physrep.2006.11.001}, 438, 237

\bibitem[\protect\citeauthoryear{Matijevič, Prša, Orosz, Welsh, Bloemen  \& Barclay}{Matijevič et~al.}{2012}]{Matijevi2012}
Matijevič G.,  Prša A.,  Orosz J.~A.,  Welsh W.~F.,  Bloemen S.,   Barclay T.,  2012, \mn@doi [The Astronomical Journal] {10.1088/0004-6256/143/5/123}, 143, 123

\bibitem[\protect\citeauthoryear{Milone}{Milone}{1968}]{milone1968peculiar}
Milone E.,  1968, Astronomical Journal, Vol. 73, p. 708-711 (1968), 73, 708

\bibitem[\protect\citeauthoryear{{Misra}, {Harikrishnan}, {Mukhopadhyay}, {Ambika}  \& {Kembhavi}}{{Misra} et~al.}{2004}]{Misra2004ApJa}
{Misra} R.,  {Harikrishnan} K.~P.,  {Mukhopadhyay} B.,  {Ambika} G.,   {Kembhavi} A.~K.,  2004, \mn@doi [\apj] {10.1086/421005}, \href {https://ui.adsabs.harvard.edu/abs/2004ApJ...609..313M} {609, 313}

\bibitem[\protect\citeauthoryear{{Misra}, {Harikrishnan}, {Ambika}  \& {Kembhavi}}{{Misra} et~al.}{2006}]{Misra2006AdSpRa}
{Misra} R.,  {Harikrishnan} K.~P.,  {Ambika} G.,   {Kembhavi} A.~K.,  2006, \mn@doi [Advances in Space Research] {10.1016/j.asr.2005.10.061}, \href {https://ui.adsabs.harvard.edu/abs/2006AdSpR..38.2897M} {38, 2897}

\bibitem[\protect\citeauthoryear{{Modak}, {Chattopadhyay}  \& {Chattopadhyay}}{{Modak} et~al.}{2022}]{Modak2022a}
{Modak} S.,  {Chattopadhyay} T.,   {Chattopadhyay} A.~K.,  2022, \mn@doi [\apss] {10.1007/s10509-022-04050-9}, \href {https://ui.adsabs.harvard.edu/abs/2022Ap&SS.367...19M} {367, 19}

\bibitem[\protect\citeauthoryear{Packard, Crutchfield, Farmer  \& Shaw}{Packard et~al.}{1980}]{packard1980geometry}
Packard N.~H.,  Crutchfield J.~P.,  Farmer J.~D.,   Shaw R.~S.,  1980, Physical review letters, 45, 712

\bibitem[\protect\citeauthoryear{Paegert, Stassun  \& Burger}{Paegert et~al.}{2014}]{Paegert2014a}
Paegert M.,  Stassun K.~G.,   Burger D.~M.,  2014, \mn@doi [The Astronomical Journal] {10.1088/0004-6256/148/2/31}, 148, 31

\bibitem[\protect\citeauthoryear{Phillipson, Boyd, Smale  \& Vogeley}{Phillipson et~al.}{2020}]{Phillipson2020}
Phillipson R.~A.,  Boyd P.~T.,  Smale A.~P.,   Vogeley M.~S.,  2020, \mn@doi [Monthly Notices of the Royal Astronomical Society] {10.1093/mnras/staa2069}, 497, 3418

\bibitem[\protect\citeauthoryear{{Pr{\v{s}}a} et~al.,}{{Pr{\v{s}}a} et~al.}{2011}]{Prsa2011a}
{Pr{\v{s}}a} A.,  et~al., 2011, \mn@doi [\aj] {10.1088/0004-6256/141/3/83}, \href {https://ui.adsabs.harvard.edu/abs/2011AJ....141...83P} {141, 83}

\bibitem[\protect\citeauthoryear{Qian}{Qian}{2001}]{qian2001orbital}
\textcolor{red}{Qian S.,  2001, Monthly Notices of the Royal Astronomical Society, 328, 914}

\bibitem[\protect\citeauthoryear{Qian}{Qian}{2003}]{qian2003overcontact}
\textcolor{red}{Qian S.,  2003, Monthly Notices of the Royal Astronomical Society, 342, 1260}

\bibitem[\protect\citeauthoryear{{Sarro}, {S{\'a}nchez-Fern{\'a}ndez}  \& {Gim{\'e}nez}}{{Sarro} et~al.}{2006}]{Sarro2006a}
{Sarro} L.~M.,  {S{\'a}nchez-Fern{\'a}ndez} C.,   {Gim{\'e}nez} {\'A}.,  2006, \mn@doi [\aap] {10.1051/0004-6361:20052830}, \href {https://ui.adsabs.harvard.edu/abs/2006A&A...446..395S} {446, 395}

\bibitem[\protect\citeauthoryear{Schafer}{Schafer}{2011}]{schafer2011savitzky}
Schafer R.~W.,  2011, IEEE Signal processing magazine, 28, 111

\bibitem[\protect\citeauthoryear{Schinkel, Marwan, Dimigen  \& Kurths}{Schinkel et~al.}{2009}]{schinkel2009confidence}
\textcolor{red}{Schinkel S.,  Marwan N.,  Dimigen O.,   Kurths J.,  2009, Physics Letters A, 373, 2245}

\bibitem[\protect\citeauthoryear{Schreiber \& Schmitz}{Schreiber \& Schmitz}{1996}]{schreiber1996improved}
\textcolor{red}{Schreiber T.,  Schmitz A.,  1996, Physical review letters, 77, 635}

\bibitem[\protect\citeauthoryear{Shore}{Shore}{2003}]{SHORE200377}
Shore S.~N.,  2003, in Meyers R.~A.,  ed., , Encyclopedia of Physical Science and Technology (Third Edition), third edition edn, Academic Press, New York, pp 77--92, \mn@doi{https://doi.org/10.1016/B0-12-227410-5/00052-1}

\bibitem[\protect\citeauthoryear{{Shu}}{{Shu}}{1982}]{Shu1982book}
{Shu} F.~H.,  1982, {The Physical Universe}

\bibitem[\protect\citeauthoryear{{Slawson} et~al.,}{{Slawson} et~al.}{2011}]{Slawson2011a}
{Slawson} R.~W.,  et~al., 2011, \mn@doi [\aj] {10.1088/0004-6256/142/5/160}, \href {https://ui.adsabs.harvard.edu/abs/2011AJ....142..160S} {142, 160}

\bibitem[\protect\citeauthoryear{Sriram, Malu, Choi  \& Rao}{Sriram et~al.}{2017}]{sriram2017study}
\textcolor{red}{Sriram K.,  Malu S.,  Choi C.,   Rao P.~V.,  2017, The Astronomical Journal, 153, 231}

\bibitem[\protect\citeauthoryear{{Sukov{\'a}}, {Grzedzielski}  \& {Janiuk}}{{Sukov{\'a}} et~al.}{2016}]{Sukova2016a}
{Sukov{\'a}} P.,  {Grzedzielski} M.,   {Janiuk} A.,  2016, \mn@doi [\aap] {10.1051/0004-6361/201526692}, \href {https://ui.adsabs.harvard.edu/abs/2016A&A...586A.143S} {586, A143}

\bibitem[\protect\citeauthoryear{Takens}{Takens}{1981}]{Takens1981}
Takens F.,  1981, in Rand D.,  Young L.-S.,  eds, Dynamical Systems and Turbulence, Warwick 1980. Springer Berlin Heidelberg, Berlin, Heidelberg, pp 366--381

\bibitem[\protect\citeauthoryear{Tan, Algar, Corr{\^e}a, Small, Stemler  \& Walker}{Tan et~al.}{2023}]{tan2023selecting}
Tan E.,  Algar S.,  Corr{\^e}a D.,  Small M.,  Stemler T.,   Walker D.,  2023, Chaos: An Interdisciplinary Journal of Nonlinear Science, 33

\bibitem[\protect\citeauthoryear{Theiler, Eubank, Longtin, Galdrikian  \& Farmer}{Theiler et~al.}{1992}]{theiler1992testing}
\textcolor{red}{Theiler J.,  Eubank S.,  Longtin A.,  Galdrikian B.,   Farmer J.~D.,  1992, Physica D: Nonlinear Phenomena, 58, 77}

\bibitem[\protect\citeauthoryear{Tran, Levine, Rappaport, Borkovits, Csizmadia  \& Kalomeni}{Tran et~al.}{2013}]{tran2013anticorrelated}
Tran K.,  Levine A.,  Rappaport S.,  Borkovits T.,  Csizmadia S.,   Kalomeni B.,  2013, The Astrophysical Journal, 774, 81

\bibitem[\protect\citeauthoryear{Wang, Zhang, Deng, Luo, Luo  \& Zhang}{Wang et~al.}{2015}]{wang2015flip}
Wang K.,  Zhang X.,  Deng L.,  Luo C.,  Luo Y.,   Zhang J.,  2015, The Astrophysical Journal, 805, 22

\makeatother
\end{thebibliography}








\bsp	
\label{lastpage}
\end{document}